\documentclass[useAMS,usenatbib]{mn2e}
\usepackage[]{txfonts}
\def\simless{\mathbin{\lower 3pt\hbox
{$\rlap{\raise 5pt\hbox{$\char'074$}}\mathchar"7218$}}}   
\def\simmore{\mathbin{\lower 3pt\hbox
{$\rlap{\raise 5pt\hbox{$\char'076$}}\mathchar"7218$}}}   
\newcommand{\be}{\begin{equation}}
\newcommand{\ee}{\end{equation}}
\newcommand{\eqb}{\begin{eqnarray}}
\newcommand{\eqe}{\end{eqnarray}}

\newcommand{\Eg}{E_{\gamma}}
\newcommand{\eb}{\epsilon_{\rm B}}
\newcommand{\sth}{\sigma_{\rm T}}
\newcommand{\rdiss}{r_{\rm diss}}
\newcommand{\mel}{m_{\rm e}}
\newcommand{\mpr}{m_{\rm p}}
\newcommand{\lp}{\ell_{\rm p}}
\newcommand{\tdyn}{t_{\rm dyn}}
\newcommand{\tacc}{t_{\rm acc}}
\newcommand{\tsyn}{t_{\rm syn}}
\newcommand{\tpg}{t_{p\gamma}}
\newcommand{\rpg}{r_{p \gamma}}
\newcommand{\rsyn}{r_{\rm syn}}
\newcommand{\gp}{\gamma_{\rm p}}
\usepackage{graphicx}
\usepackage{epsfig} 
\usepackage{epstopdf}
\usepackage{subfig}
\usepackage{tabularx}

\title[IceCube neutrinos and GRBs ]
{Implications of a PeV neutrino spectral cutoff in GRB models}

\author[M.P, D.G, S.D.]
{M. Petropoulou$^{1,2}$\thanks{E-mail: mpetropo@purdue.edu  (MP)},  D. Giannios$^{1}$\thanks{E-mail: dgiannio@purdue.edu  (DG)} 
and S. Dimitrakoudis$^3$\thanks{E-mail: sdimis@noa.gr (SD)}
\\ $^{1}$Department of Physics and Astronomy, Purdue University, 525 Northwestern
Avenue, West Lafayette, IN 47907, USA\\
$^{2}$Einstein Postdoctoral Fellow\\
$^{3}$Institute for Astronomy, Astrophysics, Space Applications \& Remote Sensing, National Observatory of Athens, 15 236 Penteli, GREECE}
\begin{document}
\date{Received / Accepted}
\pagerange{\pageref{firstpage}--\pageref{lastpage}} \pubyear{2014}

\maketitle

\label{firstpage}

\begin{abstract}
The recent discovery of extragalactic PeV neutrinos opens a new window 
to the exploration of cosmic-ray accelerators. 
The observed PeV neutrino flux is close to the Waxman-Bahcall upper bound 
implying that gamma-ray bursts (GRBs) may be the source of ultra-high energy cosmic rays (UHECRs). 
Starting with the assumption of the GRB-UHECR connection, we 
show using both analytical estimates and numerical simulations
that the observed neutrinos can originate at the jet as a result
of photopion interactions with the following implications: the neutrino spectra 
are predicted to have a cutoff at energy $\simless 10$~PeV; the dissipation responsible for the GRB emission and
cosmic-ray acceleration takes place at distances $\rdiss \simeq 3\times 10^{11}-3\times 10^{13}$~cm from the central engine;
the Thomson optical depth at the dissipation region is $\tau_{\rm T} \sim 1$; the jet carries a substantial fraction
of its energy in the form of Poynting flux at the dissipation region, and has a Lorentz factor $\Gamma \simeq 100-500$.
 The non-detection of ~PeV neutrinos coincident with GRBs will indicate that GRBs 
are either poor cosmic accelerators or the dissipation takes place 
at small optical depths in the jet.
\end{abstract} 
  

\section{Introduction} 
\label{intro}

Gamma-ray bursts (GRBs) are brief flashes of gamma-rays, which are believed to form when 
energy is dissipated internally in an ultrarelativistic jet flow
(see  \citealt{piran04, meszaros06}, for reviews). 
The mechanisms behind the energy release and the radiative
processes involved remain hotly debated. Both synchrotron emission in 
optically thin conditions $\tau_{\rm T}\ll 1$ \citep{katz94, reesmeszaros94, sarietal96} and dissipation at $\tau_{\rm T}\sim 1$ resulting in a 
distorted photospheric spectrum \citep{thompson94,meszarosrees00, pe'er06, giannios06, beloborodov10, giannios12} have been explored in the literature. 

GRBs are among the few known astrophysical sources powerful enough to
accelerate ultra-high energy cosmic rays (UHECRs) up to $10^{20}$~eV \citep{waxman95, vietri95}. 
The fact that cosmic-rays (CRs) at $\simmore 10^{19}$~eV are
injected at a rate similar to the observed $\gamma$-ray production 
rate from GRBs makes this association interesting \citep{wb99}. The coexistence
of CRs and $\gamma$-rays in the jet results in photopion interactions
and ultimately in the production of $\sim$~PeV neutrinos. The recent
detection of  IceCube neutrinos at $\sim 2$~PeV \citep{icecube13} with flux close to
the Waxman-Bahcall (WB) upper bound strengthens the GRB-UHECR connection,  although 
other sources, such as ultra-long GRBs, have been also suggested as good candidates (see e.g. \citealt{murase13} and references therein). 
An intriguing discovery of IceCube is a likely break or cutoff of the neutrino spectrum at $\simless$~10 PeV.

If the observed PeV neutrinos form at the GRB emitting region, 
then the neutrino spectrum carries 
important information about the 
conditions of the accelerator (see also \citealt{zhangkumar13}). 
In the simplest scenario, the GRB neutrino spectrum is expected to be
flat in $\nu F_{\nu}$ units, reflecting the injected proton spectrum. However,  
this is not always the case, since the proper treatment of other effects, 
such as multipion production and secondary photon emission, may cause deviations from the simple flat spectrum \citep{bhw12,asanomeszaros14, petro14, winter14}.
The putative break or cutoff of the neutrino 
spectrum can naturally arise from the synchrotron cooling of charged pions, muons and kaons, and in this case, the location of the break strongly constrains
the strength of the magnetic field at the source. This turns out to
set stringent constraints on the location where dissipation takes
place in the jet.

Here we assume that GRBs are the source of UHECRs and explore the 
implications from the presence of a break in the neutrinos. 
The present study is structured as follows: in \S2 we exploit current information about the high-energy neutrino spectrum and derive analytical constraints
for the dissipation distance in GRB flows. In \S3 we complement the previous analysis by numerical calculations of
GRB neutrino spectra for various parameter sets. We discuss the implications of our results on the nature of the dissipation mechanism
in \S4, and conclude with a summary in \S5.
\section{IceCube neutrinos}
\label{neutrinos}
Currently, high-energy neutrino astronomy has produced two significant observational findings, which are summarized below:
\begin{enumerate}
 \item the detection of $\sim$ PeV energy neutrinos of astrophysical origin
 \item the all-flavour neutrino flux in the range $100$ TeV - $2$ PeV is  reported to be
 $\sim 3.6 \times 10^{-8}$~ GeV cm$^{-2}$ s$^{-1}$ sr$^{-1}$ \citep{icecube13}, i.e. 
 close to the Waxman-Bahcall (WB) upper limit \citep{wb99}.
\end{enumerate}
Moreover, there is an indication of a spectral cutoff or softening of the neutrino
spectrum between $2-10$~PeV \citep{icecube13, aartsen14}, whose importance for the
GRB physics will be discussed in the next paragraphs. 

\subsection{Model description}
Let us consider a GRB flow of kinetic (isotropic equivalent) luminosity
$L_{\rm k}$ and bulk Lorentz factor $\Gamma$.
When the jet reaches a distance $r_{\rm diss}$ a substantial fraction
of its luminosity is dissipated internally, either through shocks (e.g. \citealt{reesmeszaros94}) 
or magnetic reconnection  (e.g. \citealt{spruitetal01}). Here, the distance $\rdiss$ is treated as a free parameter to be 
constrained by neutrino observations.
Part of the dissipated energy results in the prompt
GRB radiation $L_{\gamma}=\epsilon_{\gamma}L_{\rm k}$, where observations
indicate that $\epsilon_{\gamma}$ is of order unity. The radiation mechanism itself still remains a subject of debate with
synchrotron radiation of co-accelarated electrons \citep{katz94, reesmeszaros94, chiangdermer99} and emission from the GRB photosphere \citep{goodman86, meszarosrees00, giannios06, giannios12} being usually advocated.
For the purposes of the present study, however,  it is sufficient
to assume that the gamma-ray emission is produced
at or  close to the region where cosmic rays are accelerated.  We refer to this region as the `dissipation region'. The gamma-ray compactness can be then defined as
\eqb
\ell_{\gamma} = \frac{\sth L_{\gamma}}{4 \pi \rdiss \Gamma^3 \mel c^3},
\label{lg}
\eqe
while the spectrum is approximated by a Band function \citep{band09} with  an observed peak energy $E_{\gamma}^{\rm obs}$ in the range
0.3-0.6 MeV for high luminosity GRBs, i.e. bursts with isotropic luminosities $10^{51}-10^{52}$~erg/s \citep{ghirlanda05}. For 
for the high and low energy slopes of the Band spectrum we adopt  as indicative values $\alpha=1$ and $\beta=2.2$, respectively.

Acceleration of hadrons into a power-law form of $dN_{\rm p}/dE \propto E^{-p}$ with $p\simeq 2$ that extends to
energies as high as $10^{19}-10^{20}$~eV is likely to occur both in shocks \citep{vietri95, waxman95} and magnetic reconnection \citep{giannios10} scenarios, thus making GRB sources
potential UHECR accelerators.
The high energy cutoff of the proton distribution is determined by the balance
between the acceleration and radiation mechanisms that act respectively as energy gain and loss processes.
The accelerated protons are subsequently injected with luminosity $L_{\rm p}$ into the cooling zone, where we assume that
they are affected only by energy loss processes. In general, the proton luminosity is a multiple of the
gamma-ray luminosity, i.e. $L_{\rm p}=\eta L_{\gamma}$ with $\eta \simeq 1-10$. Here, we adopt $\eta=1$,
since values  as high as 10 may lead to significant 
distortions of the GRB electronmagnetic (EM) spectrum because of hadronic initiated EM cascades.
\citep{petro14}.
We further assume that in both the acceleration and cooling regions, 
the magnetic field and the gamma-ray photon field are the same. 
We note that our model treats in detail all the physical processes that take place only in the cooling
region in contrast to two-zone models where the emission from both regions is taken
into account \citep{reynoso14, winter14}.
\subsection{Neutrino energy and fluence}
In GRBs the local radiation field is generally strong and UHE protons may lose a significant fraction of their
energy through photopion ($p\gamma$) interactions with the GRB photons \citep{waxmanbahcall97, rachenmeszaros98}.
Here we summarize why neutrinos of $\sim$~PeV energy are expected from these interactions and connect the neutrino 
fluence to the properties of the GRB flow.

The energy threshold condition for $p\gamma$ interactions
with GRB photons at the peak of the Band spectrum can be written as
\eqb
E_{\rm p}^{\rm obs} > E_{\rm p, th}^{\rm obs} = 1.2 \times 10^{15} \ \Gamma_2^2 \left(\frac{2.5}{1+z}\right) \left(\frac{0.5 {\rm MeV}}{E_{\gamma}^{\rm obs}}\right) \ {\rm eV}
\label{Epth}
\eqe
where $z$ is the redshift of the burst and $\Gamma_2=\Gamma/100$. From this point on and throughout the text we will adopt the notation 
$Q_{x}=Q/10^{x}$ in cgs units for dimensional quantities, unless stated otherwise. We also  drop the ``obs'' qualification in order to simplify the notation. 
Charged pions that are produced with energy $E_{\pi} = \kappa_{p\gamma} E_{\rm p, th}$, where $\kappa_{p\gamma} \simeq 0.2$ is the 
inelasticity for interactions close to the threshold, decay into lighter particles after $\tau_{\pi^\pm}\simeq 2.8\times 10^{-8}$~s and
give (anti)neutrinos\footnote{Throughout the text we refer to both neutrinos and antineutrinos commonly as neutrinos.}  either directly through $\pi^{+} \rightarrow \mu^{+} + \nu_{\mu}$  or indirectly through $\mu^{+} \rightarrow e^{+} + \nu_{\rm e}+ \bar{\nu}_{\mu}$.
Eventually, each neutrino carries approximately $1/4$ of the energy of the parent pion
\eqb
E_{\nu, \rm th} \simeq 2.4\times10^{13} \ \Gamma_2^{2} \left(\frac{2.5}{1+z}\right)^2 \left(\frac{0.5 {\rm MeV}}{\Eg}\right) {\rm eV},
\label{Ev}
\eqe
where we assumed that the energy of the pion before it decays has not been reduced with respect to its
energy at production and the subscript `th' is used to remind the energy of the initial proton (see eq.~(\ref{Epth})).
Thus, the production of $\sim$ PeV neutrinos is a natural prediction of GRB 
models that advocate proton acceleration to UHE -- see point (i) in \S\ref{neutrinos}. 
The energy given by eq.~(\ref{Ev}) is related to the low-energy break of the neutrino spectrum
expected from GRBs (e.g. \citealt{guetta04, zhangkumar13}). However, the presence of this break is not always
clear, as it depends on the shape of the overall neutrino spectrum, which in turn is affected by other
parameters, such as the optical depth for $p\gamma$ interactions (see also \S\ref{results}, for detailed numerical results).

The WB upper bound \citep{wb99} is an upper limit on 
the expected neutrino fluence from GRBs under the assumption that these
are indeed the sources of UHECR acceleration and are also optically thin
to $p\gamma$ interactions. Ever since the original calculation, the WB upper bound
consists a benchmark value for GRB models and for the sake of completeness we briefly outline
the calculation. 
The local injection rate of UHECRs in the range $10^{19}-10^{21}$~eV is $\sim 10^{44}$~erg Mpc$^{-3}$ yr$^{-1}$ (e.g. \citealt{cholishooper12}).
For a flat injection spectrum ($p\simeq 2$) this corresponds to an injection rate of $\sim 5\times 10^{44}$~erg/s 
at the source in the range $\simeq 10^{11}-10^{21}$~eV. This is to be compared to the local injection rate
of $\gamma$-rays from GRBs, namely  $\simeq 4\times
10^{44}$~erg Mpc$^{-3}$ yr$^{-1}$, as found by e.g. integrating the GRB lunimosity function of \cite{wandermanpiran10}.
If $f_{\pi}$ denotes the fraction of the energy lost by protons through $p\gamma$ interactions, then 
the resulting all-flavour neutrino flux is given by \citep{cholishooper12}
\eqb
E_\nu^2\Phi_\nu \sim 6 \times 10^{-8} f_{\pi} \frac{\xi_{\rm z}}{3} \quad  {\rm GeV cm^{-2} sr^{-1} s^{-1}}, 
\label{fluence}
\eqe
where $\xi_{\rm z} \sim 3$ accounts for the redshift evolution of the source, 
which is assumed to track the star formation rate (e.g.\citealt{waxman13}).
Assuming that all the observed neutrino flux of $(3.6 \pm 1.2)
  \times 10^{-8}$ GeV cm$^{-2}$ sr$^{-1}$ s$^{-1}$ originates from
 typical GRBs, eq.~(\ref{fluence}) implies  that $f_{\pi}\simeq 0.5-1$.

One can envision cases with $f_\pi\ll 1$ or $f_\pi \gg 1$, which, however, we do not favour.
On the one hand, one could think of a scenario where GRB jets inject more energy to non-thermal hadrons than to gamma-rays, i.e. $L_{\rm p}/L_{\gamma} \gg 1$. 
In this case, the observed neutrino flux would imply $f_{\pi} \ll 1$. However, as we will show in \S\ref{results}, the resulting neutrino 
spectra for $f_\pi \ll 1$ cannot account for the observed spectral shape in the range 100~TeV-2~PeV. In addition to this, cases with $L_{\rm p}/L_{\gamma} \gg 1$
may prove to be problematic for other reasons, such as the hadronic dominance in the GRB photon spectra \citep{asanomeszaros14, petro14}. 
On the other hand, optically thick (to $p\gamma$ interactions) scenarios where $f_\pi\gg 1$ are most likely overuled because of two reasons:
even if $L_{\gamma}\simeq L_{\rm p}$, the produced neutrino flux exceeds the observed value and in such conditions 
cosmic ray acceleration to $>10^{19}$~eV is unlikely to take place (see next section).

The previous discussion relies on the assumption that all the
  IceCube neutrino flux originates from typical GRB sources. If only a
  fraction of the PeV neutrino flux turns out to come from
  GRBs, one can only set an upper limit to $f_{\pi}\lesssim 0.5$. In the
  following, we focus on the $f_{\pi}\simeq 0.5-1$ limit but explore
  other values for $f_{\pi}$ as well.
%
%

\subsection{Constraints on the dissipation region}
We use the following basic arguments in order to put constraints on the distance of the dissipation region:
\begin{itemize}
\item  the observed neutrino flux implies that $f_\pi \simeq 0.5-1$,
 \item the acceleration of protons to UHE $\sim 10^{20}$~eV should not be hampered by cooling processes in the acceleration region.
\end{itemize}

We start with the first argument and express the fraction $f_{\pi}$ in terms of GRB observables, such
as the gamma-ray luminosity and peak energy, and of the two main unknowns in GRB models, namely the bulk Lorentz factor and the dissipation distance.
The energy loss timescale of protons because of  $p\gamma$ interactions 
with gamma-ray photons is found to be constant \citep{waxmanbahcall97, petro14} for protons having energy above $E_{\rm p, th}^{\rm obs}$ (see eq.~(\ref{Epth})). This
can be written as
\eqb
t_{p\gamma} \simeq 50\tdyn \left(\frac{100}{\Gamma}\right) \left(\frac{10}{\ell_{\gamma}}\right) \left(\frac{1+z}{2.5}\right)\left(\frac{\Eg}{0.5 {\rm MeV}} \right),
\label{tpg}
\eqe
where $\tdyn\simeq \rdiss/c\Gamma$. 
The ratio $\tdyn/\tpg$ is usually defined as $f_{\pi}$ and expresses the fraction of energy lost by protons
to pions within the expansion time:
\eqb
f_\pi = 1.5 \frac{\epsilon_{\gamma, 1/3} L_{\rm k, 52}}{r_{13} \Gamma_2^2}\left(\frac{2.5}{1+z} \right) \left( \frac{0.5 {\rm MeV}}{\Eg}\right),
\label{fpi}
\eqe
where we also used eq.~(\ref{lg}). This ratio can be directly related to 
the neutrino flux as long as the synchrotron cooling timescale is larger than $t_{p\gamma}$, which
is indeed the case for protons that are responsible for the $\sim$ PeV neutrino emission.
By normalizing $f_{\pi}$ to the value implied by the observed neutrino fluence, namely $f_{\pi}=0.5$,  we define a characteristic radius as
\eqb
r_\pi = 3 \times 10^{13} \left(\frac{2.5}{1+z}\right)
\left(\frac{0.5 {\rm MeV}}{\Eg}\right) \left(\frac{0.5}{f_\pi} \right) \epsilon_{\gamma, 1/3} L_{\rm k, 52} \Gamma_2^{-2}\ {\rm  cm}.
\label{r_pi}
\eqe
This serves as an upper limit for the dissipation distance, 
since for $\rdiss \gg r_\pi$  the efficiency of $p\gamma$ process drops significantly and $f_{\pi} \ll 1$ (see \S\ref{results}, for the implications on the neutrino spectra).

The second argument can be used in order to place a lower limit on the dissipation distance.
Here, we assume that the proton acceleration process operates close to the Bohm diffusion
limit, since such high acceleration rates can be achieved both in shocks
and magnetic reconnection regions \citep{giannios10}. In this case, the acceleration timescale is
$\tacc =\gp \mpr c/ e B$, where $\gp$ is the Lorentz factor of the proton and $B$ is the magnetic field strength
in the rest frame of the jet, which at a distance $r$ from the central engine is given by
\eqb
B= \left(\frac{\eb L_{\rm k}}{c} \right)^{1/2} \frac{1}{ r \Gamma}.
\label{B}
\eqe
In the above, $\eb$ denotes the ratio of the Poynting luminosity to the  jet kinetic luminosity\footnote{The total jet luminosity is then simply the sum of Poynting
and kinetic luminosities.}. The acceleration process competes
with energy loss processes, such as radiative and adiabatic cooling, and the balance between the two defines
a saturation (maximum) energy for the particles.

Radiative losses include proton synchrotron radiation and $p\gamma$ interactions.
Proton-proton (pp) collisions also result in energy losses for cosmic ray protons but are not important
for the parameter regime relevant to this study. Given that the cross section for inelastic
pp scattering of a cosmic ray proton with one of low energy is $\sigma_{\rm pp}\lesssim 10^{-25}$ cm$^2$, \citep{atlas11}, 
inelastic collisions become important at Thomson optical depths of the flow larger than $\tau_{\rm T}\gtrsim \sth/\sigma_{\rm pp}\simeq 7$; for the definition
of $\tau_{\rm T}$ see below.

The synchrotron cooling timescale for a proton $\tsyn = 6\pi \mel c \chi^3/ \sth B^2 \gp$, where  $\chi=\mpr/\mel$.
Demanding $\tacc \le \tsyn$ and using eq.~(\ref{B}) we 
find that the dissipation should occur at distances larger than 
\eqb
\rsyn = 5.5 \times 10^{13} \ \Gamma_2^{-3} \ E_{\rm p,20}^2 \epsilon_{\rm B, 1/3}^{1/2}L_{\rm k, 52}^{1/2} \ {\rm cm}
\label{r_syn}
\eqe
in order for the acceleration process to saturate at  $E_{\rm p}^{\max}=10^{20}$~eV. 
The above apply also to $p\gamma$ interactions that may overtake synchrotron losses
for high gamma-ray compactnesses. Using eq.~(\ref{tpg}) we find that the condition $\tacc \le \tpg$ is equivalent
to $\rdiss \ge \rpg$, where the latter is given by
\eqb
\rpg = 3\times 10^{12} \ \Gamma_2^{-1}  E_{\rm p,20} 
\epsilon_{\gamma, 1/3} \epsilon_{\rm B, 1/3}^{-1/2} L_{\rm k, 52}^{1/2} \left(\frac{2.5}{1+z} \right) \left(\frac{0.5 {\rm MeV}}{\Eg} \right)  \ {\rm cm}.
\label{rpg}
\eqe
Combining eqs.~(\ref{r_syn}) and (\ref{rpg}) we find that synchrotron losses dominate over $p\gamma$ losses, unless
$\Gamma$ exceeds 
\eqb
\Gamma > 430 \ E_{\rm p, 20}^{1/2} \left(\frac{\epsilon_{\rm B, 1/3}}{\epsilon_{\gamma, 1/3}} \right)^{1/2} \left(\frac{1+z}{2.5}\right)^{1/2}
\left(\frac{\Eg}{0.5 \ {\rm MeV}}  \right)^{1/2}.
\eqe
Finally, the acceleration mechanism competes with the expansion timescale of the
flow. However, the condition $\tacc \le \tdyn$  sets a weak constraint on the bulk Lorentz factor, i.e.
$\Gamma \gtrsim 10^3 \epsilon_{\rm 1/3}^{1/2} L_{\rm k, 52}^{1/2} E_{\rm p, 20}^{-1}$ and we will not
consider it any further.
\begin{figure}
\centering
\includegraphics[width=8cm, height=8cm]{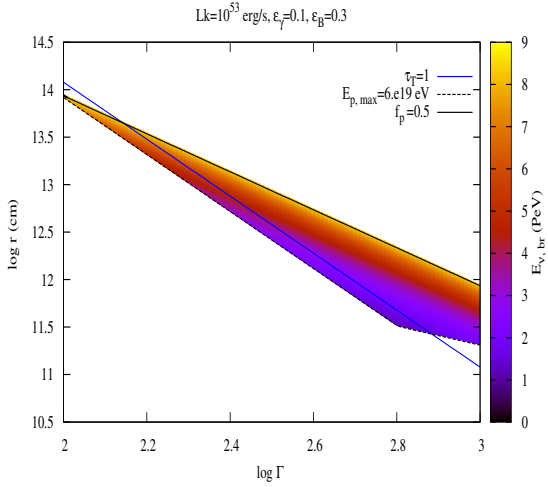}
\caption[] {The $r-\Gamma$ plane for $L_{\rm k}=10^{53}$~erg/s, $\epsilon_{\gamma}=0.1$ and $\eb=0.3$ along with
the characteristic radii $r_\pi$ (solid line), $\max(\rpg, \rsyn)$ (dashed line) and $r_{\rm ph}$ (thick blue line).
The region enclosed by the solid and dashed lines corresponds to $\rdiss$ that satisfies both the neutrino fluence and UHECR acceleration
constraints. The color coding of this region indicates the observed energy of the expected spectral cutoff in the neutrino spectrum.
The favored parameter space surrounds the Thomson photosphere ($\tau_{\rm T}=1$ line).
\label{fig2}}
\end{figure}

Combining all the above we can constrain the distance of the dissipation region between  $r_\pi$ and $\max\left(\rsyn, \rpg \right)$.
This is examplified in Fig.~\ref{fig2} 
for $L_{\rm k}=10^{53}$~erg/s, $\eb=0.3$, $\epsilon_{\gamma}=0.1$.
In addition to the above constraints we overplotted (blue thick line) for comparison reasons the `photospheric' radius ($r_{\rm ph}$),
i.e. the locus of points on the $r-\Gamma$ plane that correspond
to $\tau_{\rm T}=1$. Since the optical depth of the jet as function of distance is $\tau_{\rm T}=n_{e^{\pm}}\sth r/\Gamma=\sth L_{\rm k}/4\pi r\Gamma^3 m_p c^3$
(e.g. \citealt{meszarosrees00, giannios12}) the location of the Thomson photosphere can be written as
\eqb
r_{\rm ph} = 1.2 \times 10^{13} \frac{L_{\rm k,52}}{\Gamma_2^3} \ {\rm cm}.
\label{rph}
\eqe

Finally, the color coding indicates the {\sl predicted}  
spectral cutoff energy of the observed neutrino spectrum because of synchrotron pion cooling. 
This characteristic neutrino energy, which we will call  `break' energy from this point on, 
is given by
\eqb
E_{\nu, \rm br} \simeq 9  \left(\frac{0.5}{f_\pi} \right)
\epsilon_{\rm B, 1/3}^{-1/2} \epsilon_{\gamma, 1/3} L_{\rm k, 52}^{1/2} 
\left(\frac{2.5}{1+z}\right)^2 \left(\frac{0.5 {\rm MeV}}{\Eg} \right) \ {\rm  PeV}.
\label{ebr}
\eqe
For the above derivation we used eqs.~(\ref{r_pi}) and (\ref{B}) and the fact that $E_{\nu, \rm br}\simeq 0.25 E_{\pi, \rm c}$ where
$E_{\pi, \rm c}= \Gamma \gamma_{\pi, \rm c} m_\pi c^2$. Here $\gamma_{\pi, \rm c}$ is
the Lorentz factor
of a pion that cools because of synchrotron radiation before it decays, i.e.
\eqb
\gamma_{\pi, \rm c}=\sqrt{\frac{6\pi \mel c \chi_\pi^3}{\sth B^2 \tau_{\pi^{\pm}}}},
\label{gpic}
\eqe 
where $\chi_\pi=m_\pi/\mel$.
Figure~\ref{fig2} reveals a few things about the GRB source that are
worth commenting on: 
\begin{itemize}
 \item the dissipation takes place close to the Thomson photosphere;
 \item the constraint imposed by the observed neutrino fluence ($f_{\pi} \simeq 0.5-1$) sets an upper limit
 on the observed spectral cutoff of the neutrino spectrum, which is $\sim 10$~PeV as indicated by the color bar;
  \item the bulk Lorentz factor is less constrained as it ranges between 100 and 1000.
 \end{itemize}
If, however, GRBs prove to be only subdominant sources of the observed PeV neutrino flux, Fig.~\ref{fig2}
should be interpreted as follows: the dissipation region is placed at larger distances from the central
engine (white colored region above the $f_\pi=0.5$ line), the neutrino spectrum extends at energies above 10~PeV, while
the Lorentz factor still remains the less constrained parameter.

\begin{figure}
\centering
\includegraphics[width=8.cm, height=7.cm]{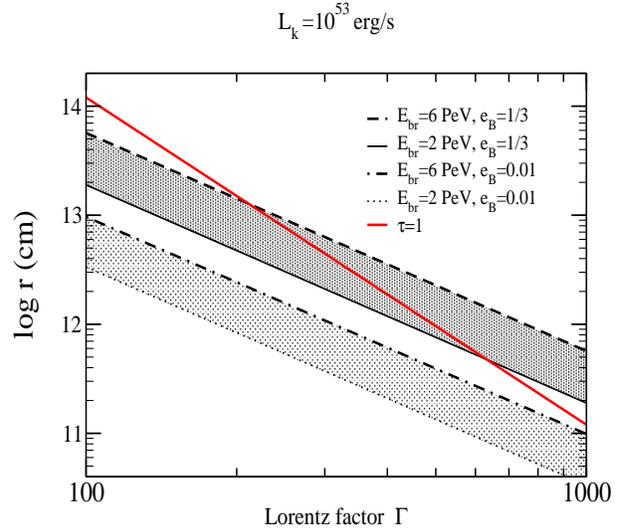}
\caption[] {The dissipation radius as function of $\Gamma$ for a break
  in the neutrino spectrum at 2 and 6 PeV and two values of $\eb$. For comparison the
  photospheric radius $r_{\rm ph}$ is also shown in red. Other parameters used are: $L_{\rm k}=10^{53}$~erg/s and $\epsilon_{\gamma}=0.1$.
  For $\eb=0.01$ dissipation takes place at $\tau_{\rm T} \gg 1$ where UHECR acceleration is less likely.
\label{fig6}}
\end{figure}

\subsection{Dependence on $E_{\nu, \rm br}$}
At the moment there is only evidence for a spectral cutoff of the neutrino spectrum between 2 and 10 PeV.
If this is confirmed, then the constraints
shown in Fig.~\ref{fig2} allow us to build a consistent picture where the dissipation
of energy occurs at such distances that favor both UHECR acceleration and
neutrino emission with flux values close to the observed one.
However, a future detection of neutrino events with a flat spectrum (in $\nu F_{\nu}$) units extending above 10 PeV
would have some interesting implications which  we will discuss in \S\ref{results} with detailed examples.

Here we keep the break energy of the neutrino spectrum  as a free parameter and we
express the various quantities introduced in \S\ref{neutrinos} in terms of $x_{\nu, \rm br} = E_{\nu, \rm br}/ 1$~PeV. 
For the magnetic field strength we find 
 \eqb
 B=10^6  \frac{\Gamma_2}{x_{\nu, \rm br}} \ {\rm G},
 \label{B-Ebr}
 \eqe
 where we used that $E_{\nu, \rm br}\simeq 0.25 E_{\pi, \rm c}$ and eq.~(\ref{gpic}). 
 Combining eqs.~(\ref{B}) and (\ref{B-Ebr}) the dissipation distance is written as
 \eqb
 r_{\rm diss}=3\times 10^{12} \frac{\epsilon_{\rm B,1/3}^{1/2}L_{52}^{1/2} x_{\nu, \rm br}}{\Gamma_2^2} \ {\rm cm}.
 \label{rdiss-Ebr}
 \eqe
Thus, if the high energy neutrino spectrum extends above a few PeV, the dissipation region should be placed
at larger distances ($\rdiss \gg r_{\rm ph}$), simply because the magnetic field is smaller further out from the central engine. 
This is also reflected at the inverse proportional dependence of $\tau_{\rdiss}$ on $x_{\nu, \rm br}$:
 \eqb
 \tau_{\rm diss}=4 \frac{L_{\rm k, 52}^{1/2}}{\Gamma_2\epsilon_{\rm B,1/3}^{1/2} x_{\nu, \rm br}},
 \label{tdiss}
 \eqe
where we used the definition $\tau_{\rm T}=\sth L_{\rm k}/4\pi r\Gamma^3 m_e c^3$ and eq.~(\ref{rdiss-Ebr}). 
Finally, 
the gamma-ray compactness is written as
 \eqb
\ell_{\gamma}=1.7\times10^3 \epsilon_{\gamma,1/3} L_{\rm k, 52}^{1/2}\Gamma_2^{-1}\epsilon_{\rm B,1/3}^{-1/2} x_{\nu, \rm br}^{-1},
\eqe 
where we used eqs.~(\ref{lg}) and (\ref{rdiss-Ebr}).  Large values of $x_{\nu, \rm br}$ correspond to a small
gamma-ray compactness, which further implies a decrease in the neutrino flux, 
given a fixed value of the ratio $L_{\rm p}/L_{\gamma}$ -- for the relation between
$\ell_{\gamma}$ and the neutrino production efficiency see \cite{petro14}. 
Finally, by equating $t_{\rm acc}=\tsyn$ and using the expression (\ref{B-Ebr}) for the magnetic field, we estimate the maximum
energy of a proton, if limited by synchrotron losses, to be: 
\eqb
E_{\rm p, syn}^{\max}=1.8\times 10^{19}\Gamma_2^{1/2}x_{\nu, \rm br}^{1/2} \ {\rm eV}.
\label{Esyn-Ebr}
\eqe
If proton acceleration is saturated by $p\gamma$ interactions we find 
\eqb
E_{\rm p, p\gamma}^{\max}=1.2\times 10^{20}\frac{x_{\nu \rm br}}{\Gamma_2^{1/2}}\frac{\epsilon_{\rm B,1/3}}{\epsilon_{\gamma,1/3}}\ {\rm eV}
\label{Epg-Ebr},
\eqe
where we used $t_{p \gamma}=t_{\rm acc}$. Thus, the maximum proton energy is typically determined by the balance between
the acceleration and synchrotron loss rates, unless $\Gamma \sim 10^3$.

The magnetization of the burst, which is one of the basic unknowns in GRB models, was kept fixed up to this point. 
Here, we investigate the role of $\eb$
on the constraints presented in Fig.~\ref{fig2}. For this, we plot the dissipation distance given by eq.~(\ref{rdiss-Ebr}) as a function of $\Gamma$
for $\eb=0.3$ and $\eb=0.01$ -- see Fig.\ref{fig6}. For each value of $\eb$ we show the expected $\rdiss$ for two indicative values of
the neutrino spectral break energy, i.e. $x_{\nu, \rm br}=2$ and 6, while $r_{\rm ph}$ is shown with thick red line. 
The dissipation region for $\eb=0.01$ is placed well inside the GRB photosphere, where other physical processes, such as pp collisions
may prevent UHECR acceleration in the first place. 
 Note that if $x_{\nu, \rm br} \gtrsim 10$, the dissipation may still be located at regions with $\tau_T \lesssim 1$ provided that
$\eb \lesssim 10^{-3}$, though in that case acceleration of protons up to $10^{20}$~eV is unlikely (see eq.~(\ref{Epg-Ebr}).
The verification of 
a spectral cutoff in the IceCube spectrum between 2 and 10 PeV will, thus, favour  substantially magnetized GRB flows with $\eb \gtrsim 0.1$.

\section{Numerical approach}
\label{results}
In the previous section we derived strong constraints on the location of 
the dissipation region using analytical arguments. 
The detailed numerical calculations of neutrino spectra reported here
fully support this analysis and provide some additional constraints, which
come from the fact that the observed neutrino spectrum is 
modeled as $E_\nu^2 \Phi_\nu \propto E_\nu^{-s}$  with $s =0-0.3$
in the 60~TeV-3~PeV energy range \citep{aartsen14}.  
First, we present  neutrino spectra obtained for a single GRB located at a fiducial redshift $z=1.5$. Then,
we calculate the diffuse GRB neutrino emission and compare our results against
the IceCube detection.

\begin{table*}
 \caption{Parameter values used for the calculation of the neutrino spectra shown in Figs.~\ref{fig3} and \ref{fig4}.}
 \begin{tabular}{cccccccc}
  \hline
 \# & $\Gamma$ & $E_{\nu, \rm br}$ (PeV) & $B$ (G) & $\rdiss$ (cm) & $\gamma_{\rm p, \max}$ & $\ell_{\gamma}$ & $\lp$ \\
   \hline
   \hline
     \multicolumn{8}{l}{$f_\pi=0.1$}\\
   \hline
 1 & 200 & 50 & $4\times 10^4$ & $10^{14}$ & $10^9$ & 17 & 0.009\\  
 2 & 300 & 50 & $6\times 10^4$ & $5.1\times 10^{13}$ & $8\times 10^8$ & 11.3 & 0.006 \\
  3 & 500 & 50 & $10^5$ & $1.8\times 10^{13}$ & $6\times10^{8}$ & 7 & 0.004 \\
 4 & 1000 & 50 & $2\times 10^5$ & $4.5\times 10^{12}$ & $4.5\times 10^8$ & 3.4 & 0.002 \\
 \hline
   \multicolumn{8}{l}{$f_\pi=0.5$}\\
   \hline
 5  & 200 & 10 & $2\times 10^5$   & $2.2\times 10^{13}$ & $4\times10^8$ & 85 & 0.05\\
 6 & 300 & 10   & $3\times 10^5$ & $9\times 10^{12}$ & $3\times 10^8$& 57 & 0.03 \\  
 7 & 500 & 10 & $5\times 10^5$ & $4 \times 10^{12}$ & $2.5\times 10^8$& 34  & 0.02 \\ 
 8 &  1000 & 10 & $10^6$ &  $9 \times 10^{11}$ & $2\times 10^8$   & 17 & 0.01 \\ 
 
 \hline
  \multicolumn{8}{l}{$f_\pi=2.5$}\\
     \hline
 9 & 200 & 3 & $6\times10^5$ & $6.7\times 10^{12}$ & $2.2\times 10^8$ & 300 & 0.2\\  
 10 & 300 & 3 & $10^6$ & $3 \times 10^{12}$ & $2\times 10^8$& 170  & 0.1 \\     
 11& 500 & 3 & $1.7\times 10^6$ & $10^{12}$ & $1.4\times 10^8$& 113  & 0.06 \\ 
 12 &  1000 & 3 & $3\times 10^6$ &  $3 \times 10^{11}$ & $10^8$   & 57 & 0.03 \\  
 \hline
 \end{tabular}
 \label{table0}
  \end{table*}

\begin{figure*}
\def\tabularxcolumn#1{m{#1}}
\begin{tabularx}{\linewidth}{@{}cXX@{}}
\begin{tabular}{c c}
\subfloat[]{\includegraphics[width=8cm]{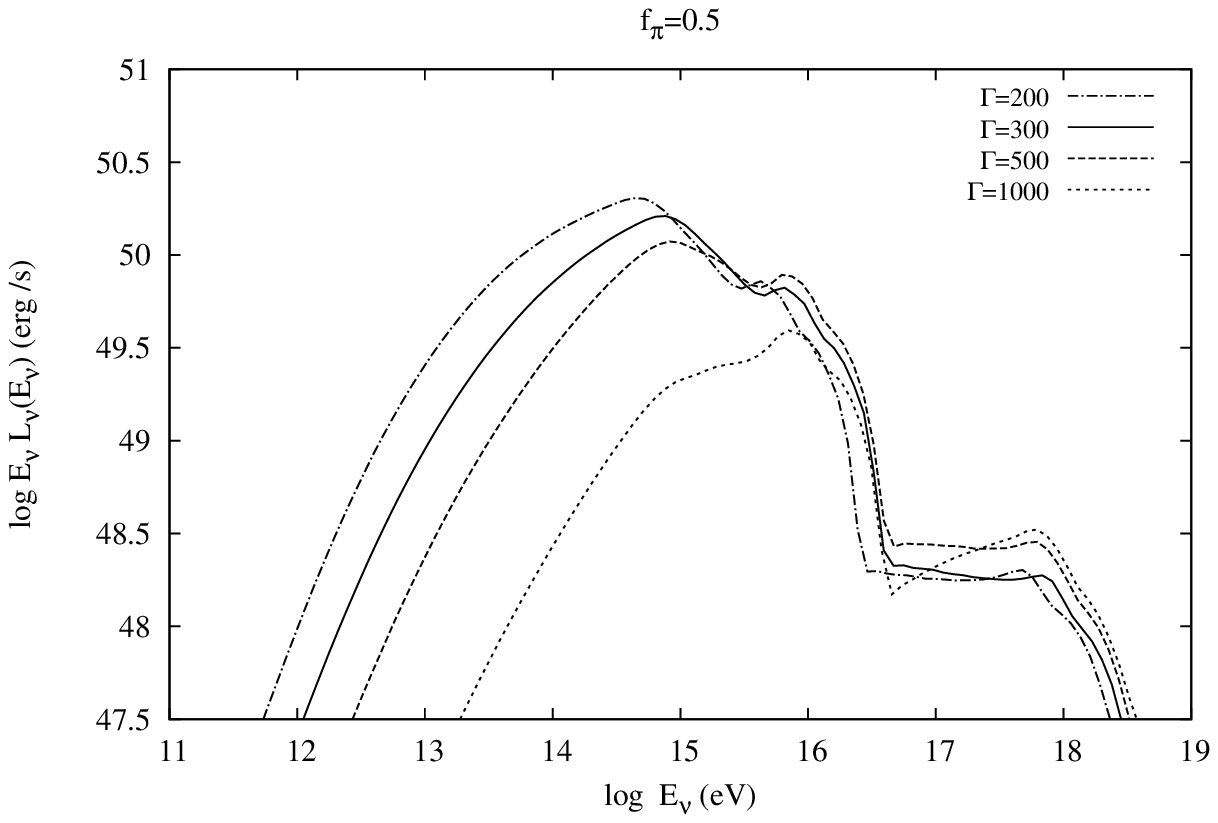}} & \subfloat[]{\includegraphics[width=8cm]{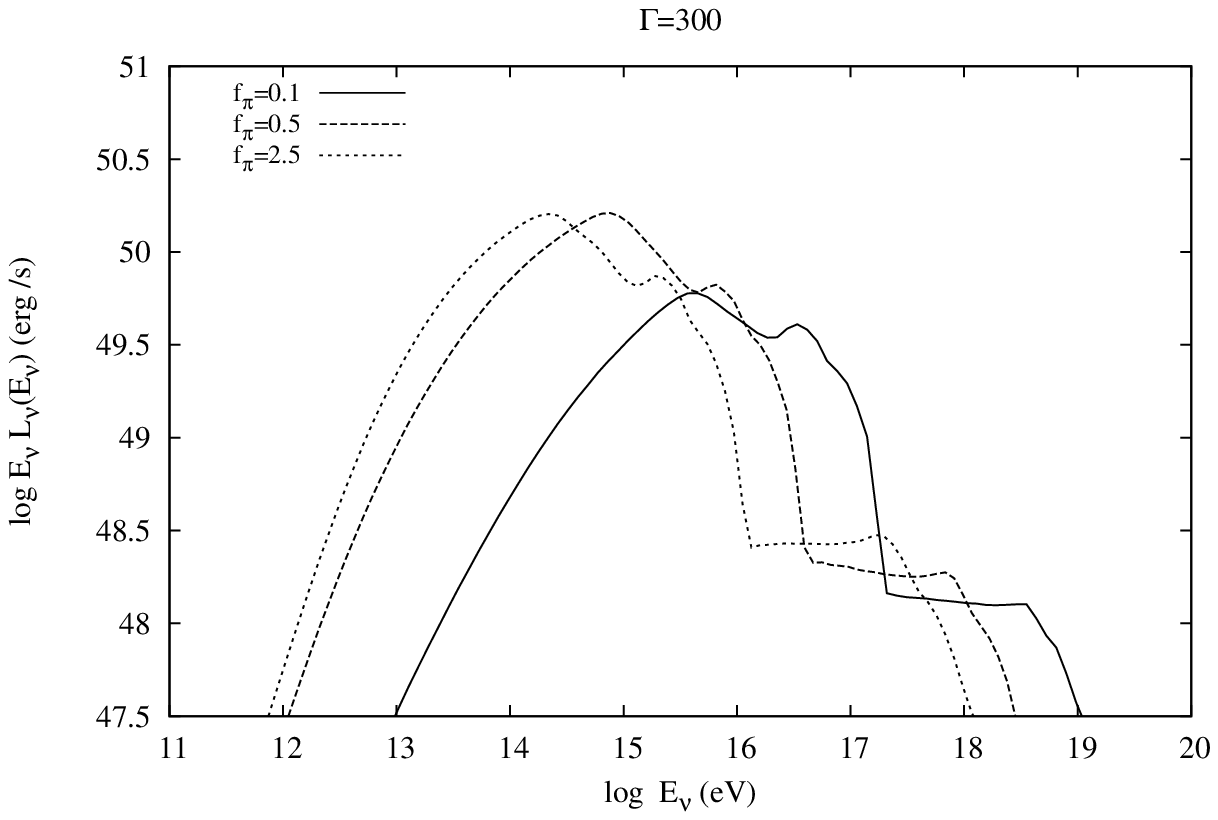}}
\end{tabular}
\end{tabularx}
\caption{All-flavour neutrino spectra for a GRB at redshift $z=1.5$. Panel (a): spectra for $f_\pi=0.5$ ($E_{\nu, \rm br}=10$~PeV) and $\Gamma=200$, 300, 500 and 1000. Panel (b):
spectra for $\Gamma=300$ and $f_\pi=0.1$, 0.5 and 2.5 that correspond to cutoff energies of 50~PeV, 10~PeV and 3~PeV, respectively.}\label{fig3}
\end{figure*}  
\subsection{Numerical code}
For this calculation we employed a kinetic equation approach, as 
described in \cite{dmpr12}-- henceforth, DMPR12. The production and loss rates of 
five stable particle species (protons, neutrons, electrons 
(including positrons), photons and neutrinos (of all flavours)) 
are tracked self-consistently with five time-dependent
equations. In addition to the processes outlined in \cite{petro14},
we now also include the effects of kaon, pion, and muon synchrotron losses,
albeit in a way that does not require the use of additional kinetic
equations. Pion, charged and neutral ($K_{\rm S}^0$ and $K_{\rm  L}^0$) kaon production rates from photo-meson 
interactions have been computed by the SOPHIA event generator 
\citep{muecke00}. For each particle energy, we calculate 
the energy lost to synchrotron radiation before it decays. The remainder
of that energy is then instantaneously transferred to the particle's
decay products, whose yields have also been computed by the 
SOPHIA event generator. Since the secondary particles from kaon
decay include pions, we first calculate charged kaon decay and then 
charged pion decay. Finally, the same process is applied to 
the resulting muons. The photons, electrons, and neutrinos resulting
from kaon, pion, and muon decay are added as production rates
to their respective kinetic equations, as are the photons from 
kaon, pion, and muon synchrotron 
radiation. Neutral kaons ($K_{\rm S}^0$ and $K_{\rm  L}^0$) and pions ($\pi^0$) are,
as in DMPR12, 
assumed to decay instantaneously, therefore directly contributing their
decay products to the kinetic equations. 

Summarizing, the numerical code as presented in DMPR12 but augmented in a way to
include pion, muon and kaon synchrotron cooling, is comparable to other
Monte Carlo (MC) codes at least in the particle physics part (e.g. \citealt{asano09, huemmer10, baerwald11,murase12,bhw12}), while
it comes with the advantage of treating time-dependent problems self-consistently (see e.g. \citealt{mastetal05, petromast12b}).
A detailed comparison of the augmented DMPR12 code with the NeuCosmA \citep{bhw12} MC code can be found
in Appendix~\ref{app1}. In the same section, we further demonstrate using the DMPR12 code the effects of other processes, 
such as neutron photopion interactions and injection of secondaries in the emission region, on the neutrino spectra.
\subsection{Numerical results} 
\label{numerical}
In total we performed twelve simulations for different values of  $\Gamma$ and $E_{\nu, \rm br}$,
while we kept fixed $L_{\rm k}=10^{53}$~erg/s, $\epsilon_{\gamma}=0.1$ and $\epsilon_{\rm B}=0.3$.
All the parameter values used in our simulations are summarized in Table~\ref{table0}.
  
Indicative neutrino spectra obtained from a single GRB at redshift $z=1.5$ are shown in Fig.~\ref{fig3}.
 Here, we plot the sum of the electron and muon
neutrino and antineutrino fluxes befor flavour mixing. For GRBs at cosmological distances, however,
the initial ratio $\nu_{\rm e}: \nu_{\mu}: \nu_{\tau}=1:2:0$ becomes $1:1:1$ because of neutrino oscillations \citep{learnedpakvasa95}. 
In this context, the neutrino spectra obtained from our simulations
are equivalent to the all-flavour  observed neutrino spectra.
The contribution of muon, charged pion and kaon decays to the total neutrino spectrum can be identified by the three
`bumps' from low to high energies in agreement to previous studies (e.g. \citealt{baerwald11}) -- see also 
Appendix~\ref{app1} for more details.

Panel (a) demonstrates the effect that $\Gamma$ has on the neutrino spectral shape for
cases with the same $f_\pi$ or equivalently $E_{\nu, \rm br}$ (see eq.~(\ref{ebr})). 
Higher values of $\Gamma$ lead to lower neutrino fluxes and harder spectra\footnote{his 
effect becomes more prominent as the source becomes more optically thin to $p\gamma$ interactions, i.e. for lower $f_\pi$. }. 
The increase of $\Gamma$  within each group of cases with the same $f_\pi$ is equivalent to stronger
magnetic fields and lower values of the gamma-ray compactness.
Higher magnetic fields cause more severe synchrotron losses to charged pions, kaons and muons, thus leading to
a decrease of the neutrino flux. Moreover, the neutrino production efficiency drops
as the source becomes less compact in gamma-rays (see also \citealt{petro14}), which
also reduces the flux, given that all other parameters are kept the same.
In all cases, the low-energy  bump of the neutrino spectrum moves to higher values for larger $\Gamma$ (see also eq.~(\ref{Ev})). For example,
the spectrum peaks at $\sim 0.8$~PeV for $E_{\nu, \rm br}=10$~PeV and $\Gamma=500$.
However, there is a qualitative change seen in the neutrino specta caused mainly by the increase of the magnetic field.
For high enough magnetic fields, e.g. $B \sim 1$~MG, 
we find that the second bump of the spectrum, which is 
related to the direct pion decay, carries most of the neutrino luminosity  (dotted line in panel (a) of Fig.~\ref{fig3}).
This result is also in agreement with the study of \cite{bhw12} (see Fig.~5 therein).
Summarizing, models that result in hard neutrino spectra, cannot account for the sub-PeV neutrino emission. 

Cases with the same value of $\Gamma$ but different cutoff energies $E_{\nu, \rm br}$ are shown in panel (b).
The neutrino flux decreases as the spectral break
moves to higher energies, which is the result of both decreasing $\ell_{\gamma}$ and $f_\pi$. 
A  future detection of neutrinos above $10$~PeV will, therefore, point towards large dissipation distances and 
low Lorentz factors, e.g. $\Gamma\lesssim 100$, in order to achieve both the flat spectral shape and the observed flux.

Having explained the basic features of the single burst neutrino emission, we proceed with the 
calculation of the diffuse neutrino flux. For this, we used as a typical duration for long GRBs $T_{\rm obs} \sim 30$~s \citep{goldstein12,gruber14} and
assumed that the GRB rate follows the star formation (SF) rate.  
In particular, we adopted the second SF model by \cite{porcianimadau01} and for the local GRB rate
we used the value derived by \cite{wandermanpiran10}, i.e. $\rho(0) \simeq 1$ Gpc$^{-3}$ yr$^{-1}$. Thus, the GRB rate as a function of redshift is
written as
\eqb
R_{\rm GRB}(z) = 23 \rho(0)\frac{e^{3.4z}}{22+e^{3.4z}} \ {\rm Gpc^{-3} yr^{-1}}.
\eqe
We obtain first from the numerical simulations the neutrino fluence as measured in the rest frame of the galaxy ($dN_{\nu}^{\rm rf}/dE_{\nu}^{\rm rf}$) and then
we calculate the diffuse neutrino flux in GeV cm$^{-2}$ s$^{-1}$ sr$^{-1}$ (see also \citealt{murasenagataki06, cholishooper12}) as
\eqb
E_{\nu}^2 \Phi_{\nu} = \frac{c}{4 \pi H_0} \int_{0}^{z_{\max}}{\rm d}z 
\frac{dN_{\nu}^{\rm rf}}{dE_{\nu}^{\rm rf}} \frac{R_{\rm GRB}(z)}{\sqrt{\Omega_\Lambda+(1+z)^3 \Omega_{\rm M}}},
\eqe
where $H_0=70$~km Mpc$^{-1}$ s$^{-1}$, $z_{\max}=9$, $\Omega_{\rm M}=0.3$ and $\Omega_{\Lambda}=0.7$ for a flat universe.

\begin{figure}
\def\tabularxcolumn#1{m{#1}}
\begin{tabularx}{\linewidth}{@{}cXX@{}}
\begin{tabular}{cc}
\subfloat[]{\includegraphics[width=8cm]{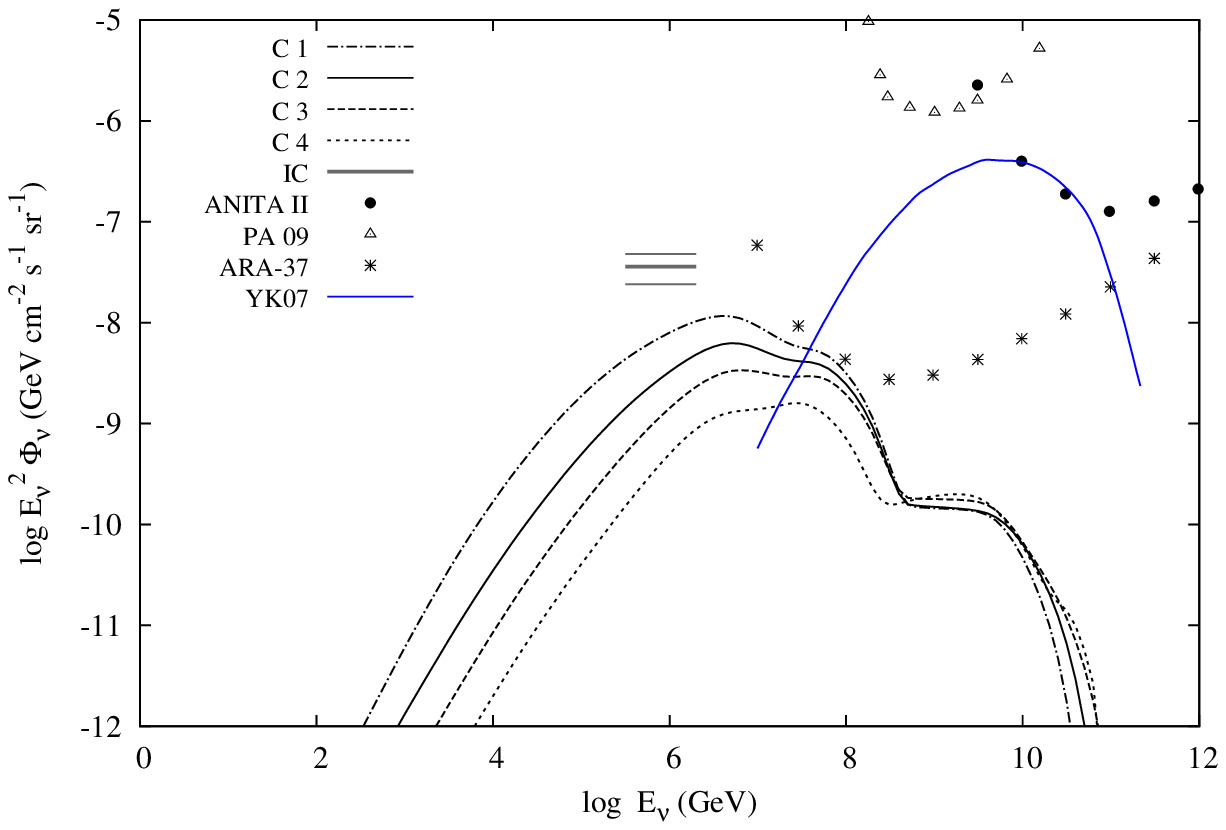}} \\
\subfloat[]{\includegraphics[width=8cm]{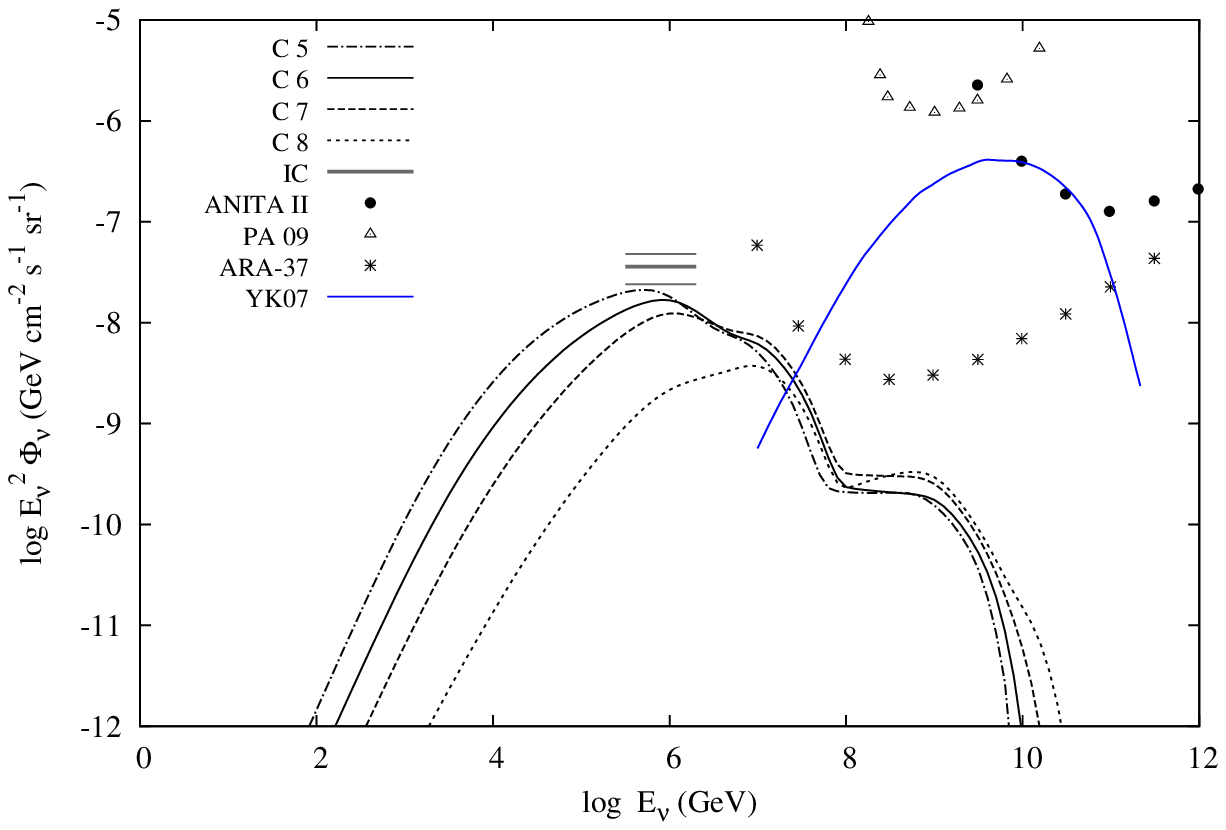}} \\
\subfloat[]{\includegraphics[width=8cm]{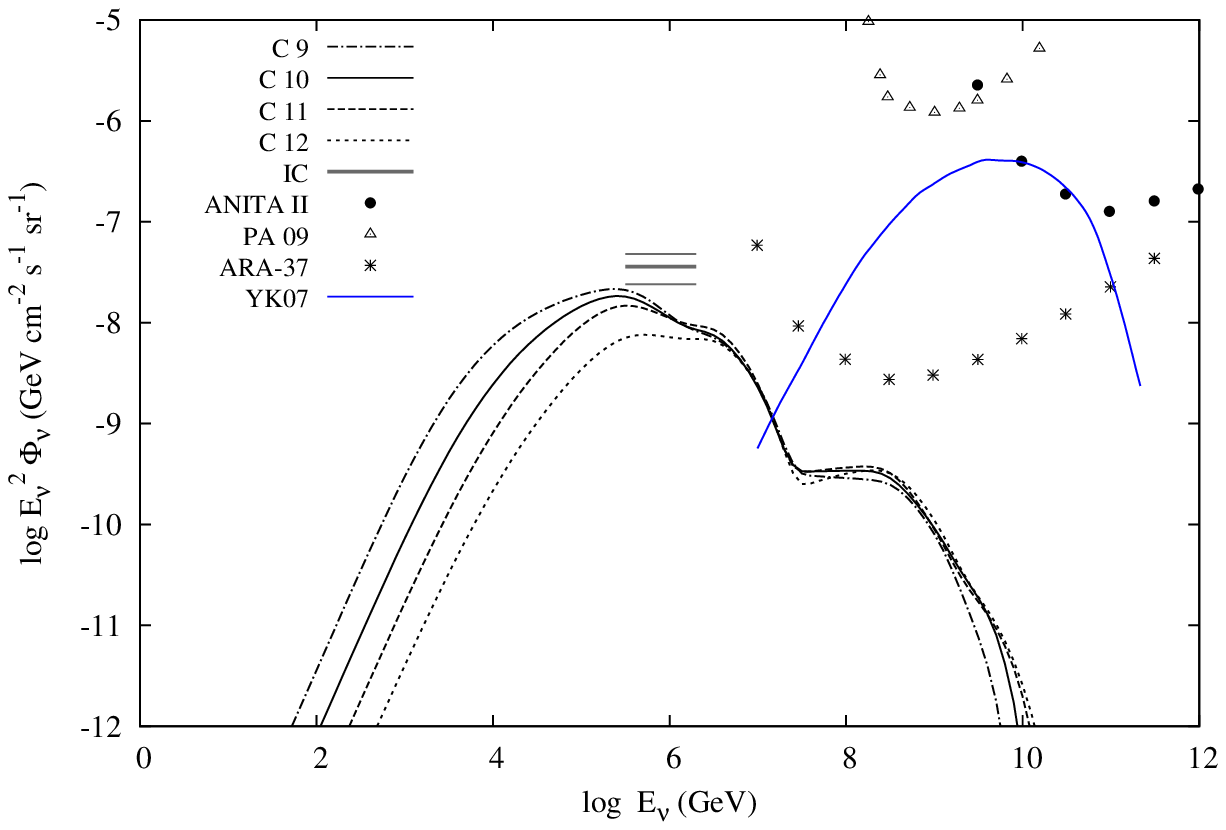}}\\
\end{tabular}
\end{tabularx}
\caption{All-flavour diffuse neutrino emission for high luminosity GRBs with typical duration $T_{\rm obs}=30$~s and $L_{\rm k}=10^{53}$~erg/s, $\eb=0.3$, $\epsilon_{\gamma}=0.1$.
Panels (a), (b) and (c) show the neutrino spectra for $E_{\nu, \rm br}=50$, 10 and 3~PeV, respectively.
In each panel, spectra are calculated for $\Gamma=200$ (dash-dotted), 300 (solid), 500 (dashed) and 1000 (dotted). 
In all panels the IceCube detection \citep{icecube13}, 
the upper limit by ANITA II \citep{ANITAII}, the upper limit on $\tau$-neutrino flux by Pierre Augere \citep{abraham08} as well as the expected 3-year sensitivity of ARA \citep{ARA12} are also shown
with thick grey lines, circles, open triangles and stars, respectively. An indicative  model for cosmogenic neutrinos by
\citet{yukselkistler07} is also plotted with a blue line.}\label{fig4}
\end{figure}

Our results are presented in Figs.~\ref{fig4}(a)-\ref{fig4}(c) along with the flux value measured by IceCube $(3.6 \pm 1.2) \times 10^{-8}$~GeV cm$^{-2}$ s$^{-1}$ sr$^{-1}$.
The upper limit of ANITA~II \citep{ANITAII}, the upper limit on $\tau$-neutrino flux by Pierre Auger \citep{abraham08}, and the expected 3-year sensitivity of ARA \citep{ARA12} are also
shown.

Cases 1-4 with spectral cutoff at $\sim 50$~PeV (panel (a)) cannot account for the observed neutrino spectra.
These are obtained for relatively large values of $\rdiss$ where the efficiency of pion production is small -- here, $f_\pi=0.1$.
One could argue that for $L_{\rm p}/L_{\gamma} \simeq 3-5$ the neutrino flux would be close to the observed value. 
Even in this case the hard spectrum below the PeV energy range would contradict the observations, except for $\Gamma <200$. However,
given that GRBs are expected to come from jets with $\Gamma \simmore  100$,
the suggestion that all bursts should be accompanied by slow jets requires fine tuning.
Cases 5-7 and 9-12  predict fluxes close\footnote{ The model derived fluxes lie close but still below the low observational 
error bar. We argue, however, that these cases can account for the IceCube observations, since
a different model for the SF rate or  $\eta=1-5$, would
result in spectra with higher fluxes and the adequate shape.} to the the IceCube measurements and
result in soft or even flat spectra, see e.g. Cases 5 and 12, respectively.
As long as GRBs are the only sources contributing to the observed PeV
flux, Cases 5-7, 9-12 favour scenarios with relatively 
small dissipation distances ($\rdiss \gtrsim r_{\rm ph}$), moderate-to-low values of $\Gamma$, 
and strong magnetic fields ($10^5-10^6$~G) -- see Table~\ref{table0}. If, however, the IceCube 
PeV neutrino-GRB connection is disfavoured (see e.g. \citealt{huemmer12, he12,liuwang13}), 
would indicate that GRBs are poor cosmic ray accelerators or, alternatively, 
large-distance dissipation scenarios would be more appropriate making Cases 1-4 more relevant.

All neutrino spectra shown in Figs.~\ref{fig4} (a)-(c) extend up to $0.1-1$~EeV.
We find that this energy range is dominated either by the exponential cutoff of the 
direct pion decay bump (Cases 1-4) or by the kaon-decay bump (Cases 5-12). 
Because of this, the expected GRB neutrino
 flux at $0.1$~EeV is only a small fraction ($1\%-10\%$) of the IceCube value.
Yet, the value $\sim 10^{-9}$~GeV cm$^{-2}$
 s$^{-1}$ sr$^{-1}$ is close to the expected sensitivity limit of next generation experiments, such as ARA \citep{ARA12}, and 
 in this respect, Cases 1-4 are promissing.
 A future detection of EeV neutrinos cannot be used, in principle, to distinguish between different GRB models mainly because of 
the contribution of cosmogenic (GZK) neutrinos at  this particular energy range \citep{beresinsky69, stecker73}. To illustrate
this, we plotted  the model of \citet{yukselkistler07} for GZK neutrinos (blue line), which predicts a
higher flux at this energy range than other GZK models (see e.g. Fig.~29 on \citealt{ARA12}). 
Although both GRBs and GZK neutrinos may contribute to this energy range, their spectra are radically different, namely soft and hard respectively (see
e.g. panel (a) in Fig.~\ref{fig4}).
This may prove to be a strong diagnostic tool, if the sensitivity of future experiments allows spectral construction.
We note also that radio-loud blazars may have a non-negligible contribution to this energy range \citep{murase14}. In any case, 
a discrimation between the various contributions seems to be necessary.
\subsection{The revised $r-\Gamma$ plane}
 \begin{figure}
 \centering
\includegraphics[width=8.5cm, height=7.5cm]{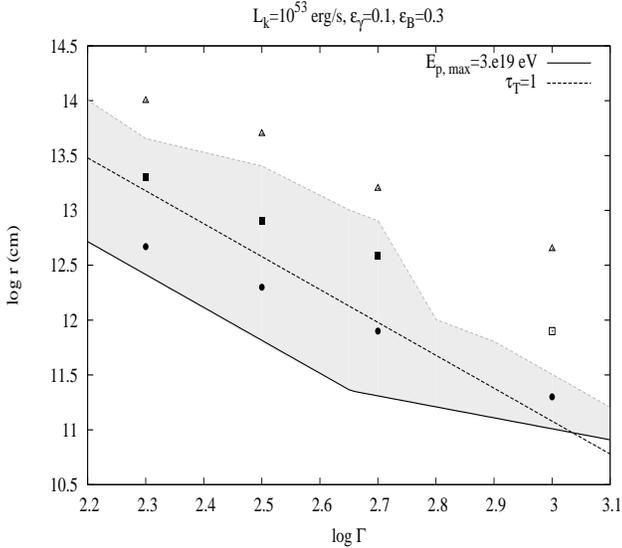}
\caption[]{Same as in Fig.~\ref{fig2} but for $E_{\rm p, \max}=3\times 10^{19}$~eV. The characteristic radii 
$\max(\rpg, \rsyn)$ and  $r_{\rm ph}$ are shown with solid and dashed black lines, respectively. 
Numerical runs corresponding to $E_{\nu, \rm br}$=50~PeV ($f_{\pi}=0.1$), 10~PeV ($f_{\pi}=0.5$) and 3~PeV ($f_{\pi}=2.5$) are shown with triangles, squares and circles, respectively.
Runs that can account for the IceCube detection are shown with filled symbols, otherwise open symbols are used. 
The grey colored region illustrates the allowed parameter space, if GRBs are the main source of PeV neutrinos. Otherwise,
the region that lies above the grey colored one is allowable.
\label{fig5}}
\end{figure}
The numerical analysis of the previous section acts complementary to the analytical approach presented in \S\ref{neutrinos}. Here, we use this additional information
in order to place even stronger constraints on the dissipation distance. The revised $r-\Gamma$ plane 
is shown in Fig.~\ref{fig5}. The characteristic radii $\max(\rpg, \rsyn)$ and  $r_{\rm ph}$
are plotted with solid and dashed lines, respectively, while the parameter sets used in \S\ref{numerical} are shown  as symbols. In particular,
sets that correspond to $E_{\nu, \rm br}=50$~PeV ($f_\pi=0.1$), 10~PeV ( $f_\pi=0.5$) and 3~PeV ($f_\pi=2.5$) are shown with triangles, squares and circles, respectively.
We use filled symbols, only if the calculated neutrino spectra can account for the current IceCube observations (see panels (a)-(c) in Fig.~\ref{fig3}); 
open symbols are used, otherwise. The final allowed parameter space (grey colored region) is truncated compared to the one shown in Fig.~\ref{fig2},
as the region for $\Gamma \gtrsim 600$ is now excluded. We caution
the reader that if the GRB-PeV neutrino
connection is disproved, then the region above the grey colored one should
be considered as allowable, thus favoring long-distance dissipation scenarios.
\section{Discussion} 
\label{discussion}
The characteristics (fluence and shape) of the high-energy neutrino spectrum expected from GRBs depends sensitively on the compactness
of the dissipation region and, for this reason, they can be used as a probe. 
In \S\ref{results} we showed how the fluence and the indication of a spectral cutoff at a few PeV 
place the dissipation region fairly close to the
Thomson photosphere (see also \citealt{murase08}). Additional information from the neutrino spectra 
favors jets with bulk Lorenz factor $\Gamma \sim 100-500$ and, hence, exclude part of the parameter space.
Interestingly, independent studies on spectral formation close and above the GRB photosphere  because of continuous energy dissipation seem
promising for the GRB emission itself (e.g. \citealt{giannios12}).
What causes therefore the dissipation and UHECR acceleration at $\tau_{\rm T} \sim 1$ in the jet? 

If the jet contains a substantial neutron component, 
energy dissipation through neutron-proton collisions is possible. 
The dissipation peaks at $\tau_{\rm T} \sim 10$ where the decoupling of the neutron and 
proton fluids takes place, and continues further out in the flow at smaller optical depths (e.g. \citealt{beloborodov10, vurm11, koersgiannios07}). 
In this picture, most of the collisions take place at mildly relativistic speeds throughout the volume of the jet, heating the flow. 
It is not obvious, however, how CRs can be accelerated to ultra-high energies in such scenario.

For high values of $\Gamma$, the dissipation distance that we infer does not differ much from the radius
of the progenitor $r_{\star}\sim 10^{11}$~cm, thus making recollimation shocks a likely culprit for the dissipation \citep{lazzati07}. However,
it is not clear whether the jet, after crossing the stellar surface, can reach a terminal $\Gamma$ of several hundreds by a distance of $\sim 10^{11}$~cm.  
This depends on several parameters, such as the magnetization of the flow and the external pressure \citep{sapountzis13}. The Lorentz factor achieved during the first
acceleration phase, which takes place inside the star, plays also a crucial role \citep{komissarovetal10}.
Moreover, even if a satisfactory dissipation mechanism operated at these small distances, the resulting neutrino spectra for $\Gamma \gtrsim 800$ would be too hard
to explain the observed spectrum.

Our analysis showed that GRB neutrino spectra are compatible with the IceCube observations for $\Gamma \simeq 100-500$ and $\rdiss \simeq 3\times 10^{11}-3\times 10^{13}$~cm.
In this range, both internal shocks and magnetic reconnection can be invoked as possible dissipation mechanisms.
Internal shocks occur at distances $r_{\rm is} \simeq 3\times 10^{11}\Gamma_2^2 \delta t_{-3}$~cm, where $\delta t$ is the observed variability timescale.
Interestingly, the same scaling $\rdiss \propto \Gamma^2$ applies also to  scenarios of magnetic reconnection.
It can be shown (see e.g. \citealt{drenkhahnspruit02}) that in a strongly magnetized jet with magnetic field reversals on a scale $L$ the reconnection distance  is
 $r_{\rm rec} \simeq 10^{12} \Gamma_2^2 L_7/ \beta_{\rm rec, -1}$~cm,
where we have assumed that the GRB central engine contains magnetic field reversals on a scale $L\simeq 10\ r_{\rm g}\simeq 10^7 L_7$~cm and that the reconnection takes place
 at the speed $\beta_{\rm rec}$.
Here, we used a conservative value for the reconnection rate. Magnetic reconnection, however, may proceed at a fairly slow rate in the collisional
$\tau_{\rm T}>1$ region and speed up at $\tau_{\rm T}\simless 1$ because of a switch from collisional to collisionless conditions \citep{mckinneyuzdensky12}. 
Although in both cases a fine tuning appears to be required so that the dissipation takes place preferentially at the inferred
distance, magnetic reconnection remains a viable mechanism for Poynting-flux dominated jets, whereas internal shocks prove to be problematic 
(e.g. \citealt{sironispitkovsky09}). With the current IceCube data our analysis points, indeed, towards strongly magnetized jets with $B\sim 10^{5}-10^6$~G.
Future verification of a spectral cutoff $\lesssim 10$~PeV will exclude GRB models with low magnetization, e.g. $\eb \lesssim 0.01$, as these
require  $\rdiss < r_{\rm ph}$.

If magnetic reconnection in Poynting-flux dominated jets proves to be the  dissipation mechanism in GRBs,
then the problem of proton acceleration in reconnection layers becomes relevant. Here, we used only rough estimates of the acceleration
and energy loss timescales and showed that UHECR can be achieved at distances close to the GRB photosphere. Recent particle-in-cell (PIC) simulations
have shown that relativistic reconnection can produce non-thermal electrons  with hard power-law spectra ($p\lesssim 2$) 
in regions with high magnetization. The electron energy is found 
to increase linearly with time close to the Bohm diffusion limit  \citep{sironispitkovsky14}. 
Although there is no definite answer to the problem of proton (ion) acceleration in relativistic reconnection, there are
indications that the acceleration process in pair plasmas and electron-ion plasmas shares many features, such
as the acceleration rate and the power-law slope (private communication with Dr.~L.~Sironi).

The above implications on the physical conditions of the GRB emission site
hold as long as the assumption of the `typical' GRBs being the sources of PeV neutrinos is still valid.
However, the non-detection of individual GRBs by IceCube so far starts putting severe constraints
on this possibility. For example, by considering the existing
IceCube limit on the neutrino flux of triggered GRBs,
this GRB population can only account for a flux of a few $10^{-9}$ GeV cm$^{-2}$
s$^{-1}$ sr$^{-1}$ (e.g. \citealt{liuwang13}). In this case, our analysis
should be translated as follows: either `typical' GRBs cannot accelerate 
CRs to UHE or the physical conditions are such as to suppress pion and neutrino production, i.e.
the grey-colored region in Fig.~\ref{fig5} would be not allowed and
the dissipation distance would have to be placed further out to the flow where $f_\pi \ll 1$ (see also
\citealt{zhangyan11}). The PeV neutrino flux could still originate from
GRBs, but from the low-luminosity class 
(e.g. \citealt{cholishooper12, murase13}), as their rate is uncertain and they would not violate the stacking limits 
derived by triggered GRBs \citep{abbasi10, abbasi11, abbasi12}.

\section{Conclusions}
\label{conclusions}

We explored the implications of the
recent  PeV astrophysical neutrino detection with IceCube on the
properties of the GRB flow.
The prompt gamma-ray emission provide the targets for photopion interactions of
protons having energy $\simmore 10^{16}$~eV. These lead to the creation of high-energy pions
that subsequently decay into $\sim$PeV neutrinos. 
In principle, the GRB neutrino spectrum is predicted to have two breaks: the (sub)PeV break is related to the
energy threshold for pion production with photons from the peak of the GRB spectrum, whereas
the second break  ($E_{\nu, \rm br}$) is related to synchrotron cooling of the parent pions.
The latter depends  on the comoving magnetic field strength 
as $B=10^6\ (\Gamma/100)/(E_{\nu, \rm br}/1~{\rm PeV})$~G.

IceCube has presented the first evidence for a cutoff in the neutrino spectrum
at energies $E \simless 10$~PeV. Given that GRB jets are expected to have $\Gamma \sim 100-1000$, 
the field strength at the emission region has to be $\sim 10^{6}$~G. This inference places the dissipation region
at a fairly compact location in the jet. We estimated the fraction of proton energy lost to pion production
within the expansion time ($f_\pi$) at such distances and found $f_\pi \sim 0.5-1$, i.e. close to the value inferred from the IceCube detection. 
Thus, an observed cutoff of the neutrino spectrum at several PeV actually implies that the neutrino flux is close to the
Waxman-Bahcall (WB) upper limit. 
We elaborate on this remark using detailed numerical simulations. In general, our numerical results 
confirmed the connection between $E_{\nu, \rm br}$
and the expected neutrino fluence,  except for cases with high Lorentz factors ($\gtrsim 800$) where the neutrino
spectra are too hard with a large curvature.

Despite the compactness of the dissipation region, protons
can be accelerated to energies up to $\sim 10^{20}$~eV
provided that the magnetic field in the jet is not very weak
$\epsilon_B\simmore 0.1$. 
On the one hand, there is evidence that the  composition of UHECRs changes
from light (protons) to heavy (Fe) for energies  $\simmore 10^{19}$~eV\citep{PA11}, and 
GRB jets may be rich in heavy nuclei, too \citep{metzger11}. 
On the other hand, observations at $\sim 10^{17}$~eV indicate a light composition for UHECRs.
Moreover, in many scenarios for UHECR acceleration, protons dominate the
composition at $\sim 100$~PeV energies, and since these are responsible for the $\sim$~PeV neutrino production, our main conclusions are left unchanged.

Summarizing, the verification of a cutoff of the neutrino spectrum at energies below a few Pev has
two profound implications for the GRB flow.
First, the jet carries a substantial fraction of its luminosity in the form of Poynting flux
and the emission region is strongly magnetized with comoving magnetic fields of $\sim 1$~MG. Second,
the dissipation of energy takes place close to the Thomson photosphere at distances $3\times 10^{11}-3\times 10^{13}$~cm.
 Unambiguous  proof of the connection between GRBs and PeV neutrinos can come
from the simultaneous detection of both high-energy signatures.   So far, however, no such detection has taken place, placing
  increasingly strict limits on the possible contribution of classical
  GRB to the ambient neutrino flux. Such a detection will not only reveal
  a strong candidate source of UHECRs but will also be a unique probe of where in the jet the dissipation takes place.
  
\section*{Acknowledgements}
We would like to thank the anonymous referee for helping improving the manuscript. We thank also 
Prof. A. Mastichiadis, Prof. Kohta Murase, Prof. W. Winter and Dr. X.~Y. Wang for useful discussions.
Support for this work was provided by NASA 
through Einstein Postdoctoral 
Fellowship grant number PF3~140113 awarded by the Chandra X-ray 
Center, which is operated by the Smithsonian Astrophysical Observatory
for NASA under contract NAS8-03060.
DG acknowledges support from the Fermi 6 cycle grant number 61122.
\bibliographystyle{mn2e} 
\bibliography{grbhadro}
\appendix
\section[]{Comparison between the DMPR12 and  NeuCosmA codes}
\label{app1}
The GRB neutrino spectrum may deviate from the often adopted Waxman-Bahcall trapezoidal spectral shape (e.g. \citealt{abbasi10}).
A more detailed treatment of the physical processes involved, such as inclusion of the multipion
production channels, may lead to more complex shapes (e.g. \citealt{baerwald11}).

Here we attempt a detailed comparison between the neutrino spectra obtained with our numerical code (DMPR12), which combines 
the physics of the SOPHIA code with the kinetic
equation approach, and those obtained with the Monte Carlo (MC) code NeuCosmA \citep{huemmer10}.
For the comparison we chose the electron and muon neutrino spectra shown in Fig.~18 of \cite{bhw12} -- henceforth BHW12.
These  are calculated at the rest frame of the emission region for $B=3\times 10^5$~G, $\gamma_{\rm p, \max}=1.1\times 10^8$, $\Gamma=10^{2.5}$, and
$z=2$. The GRB spectrum is modeled 
having a break at $\epsilon'_{\gamma, \rm br}=14.8$~keV and extending from $\epsilon'_{\min}=0.2$~eV to $\epsilon'_{\max}=300 \epsilon'_{\gamma, \rm br}$,
with photon indices below and above the peak $\alpha=1$ and $\beta=2$, respectively. 

In order to use the same assumptions as in BHW12, we modified the DMPR12 code accordingly by neglecting the:
\begin{enumerate}
 \item neutron photopion production
 \item modification of the GRB photon spectrum due to the emission of secondaries, e.g. gamma-rays from $\pi^0$ decay
 \item modification of the low-energy part of GRB spectrum because of synchrotron self-absorption
 \item modification of the high-energy ($>1$~MeV) part of the GRB spectrum due to photon-photon absorption
 \item modification of the injected proton distribution due to cooling.
 \end{enumerate}
The size of the emission region as well as the injection compactness of protons and GRB photons are necessary input quantities 
for the DMPR12 code, which
is a PDE solver, in contrast to MC codes. Given that the above quantities are not
defined in BHW12, we use the fiducial values $r_{\rm b}\simeq r_{\rm diss}/\Gamma=1.9\times 10^{11}$~cm, $\lp=10^{-2.4}$ and $\ell_{\gamma}=6.8$, and 
normalize {\sl a posteriori} the resulting neutrino spectra with respect to those in Fig.~18 of BHW12.

\begin{figure}
 \centering
\includegraphics[width=0.5\textwidth, height=8cm]{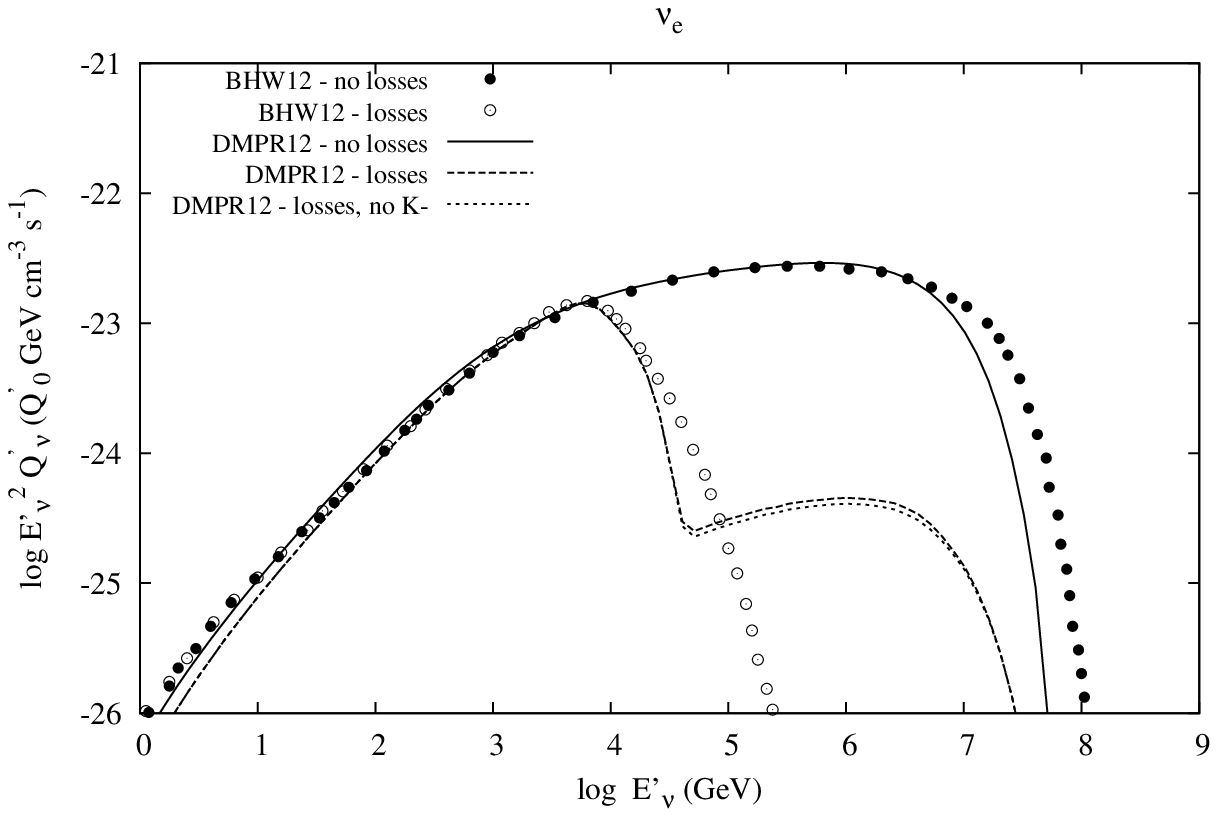}
\includegraphics[width=0.5\textwidth, height=8cm]{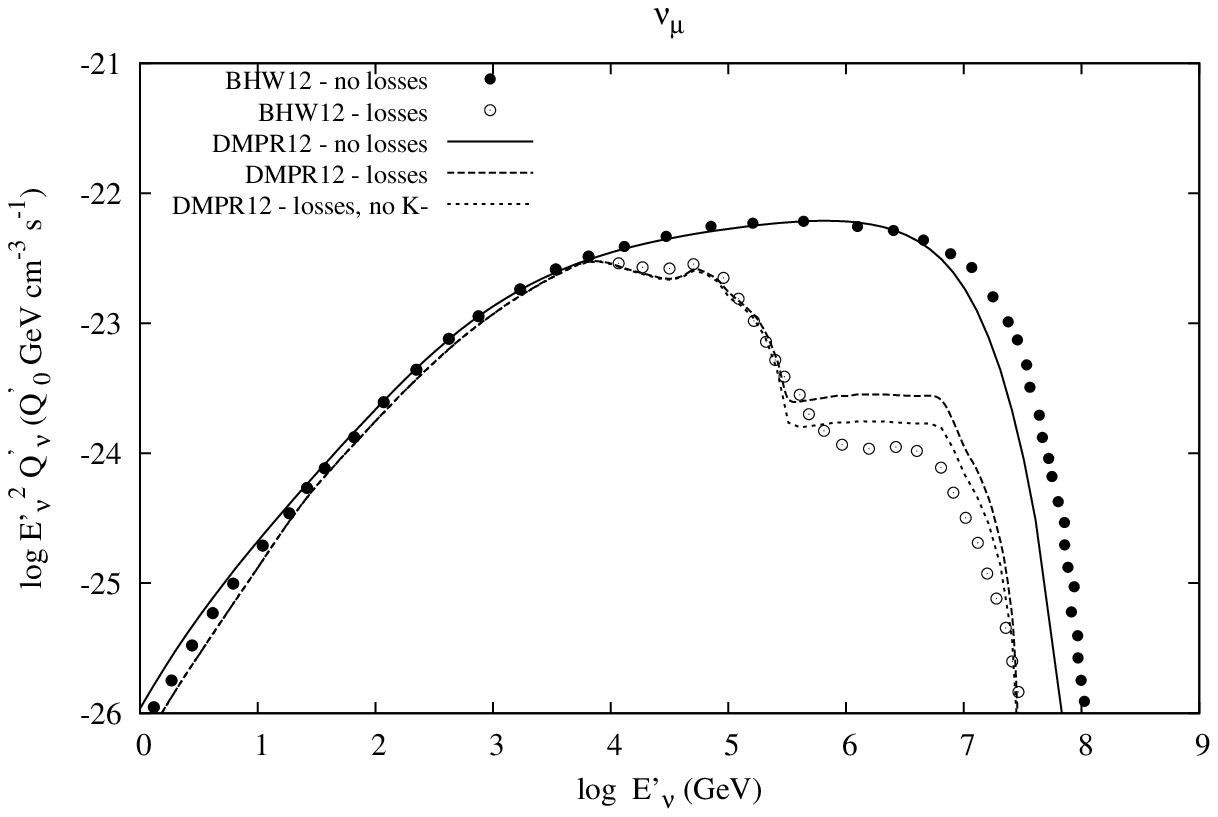}
\caption[]{Electron (top panel) and muon (bottom panel) neutrino energy spectra as measured in the comoving
frame of the emission region for the same parameters as in BHW12 -- see Table~1 therein.
The results of the NeuCosmA 2011 code are shown with 
symbols. Open and filled symbols
correspond to cases with and without synchrotron losses of secondaries, respectively. The results obtained with the DMPR12 are plotted with lines -- see legend for more details.
The dotted lines are the resulting spectra when the production of $K^{-}$ is not taken into account.
 \label{fig1-app}}
\end{figure}

The electron and muon neutrino spectra are shown in the top and bottom panels of Fig.~\ref{fig1-app}, respectively.
The neutrino spectra of BHW12 when synchrotron losses of pions, muons and kaons are taken
into account are plotted with open circles, whereas filled circles correspond the no loss case.
Our results are overplotted with solid (no losses) and dashed (with losses) lines. 
The neutrino spectra obtained by neglecting the $K^{-}$ production are also shown
with dotted lines. 

In the case where the losses of secondaries are not taken into account, we find a good agreement between the 
two codes except for a small deviation at the high-energy part of the spectra. This is caused by a difference at the high-energy cutoff
of the proton distribution, which in our case is abrupt, whereas in BHW12 is assumed to be exponential.
When synchrotron losses of secondaries are taken into account, we find that
the electron and muon neutrino spectra calculated with the DMPR12 code 
are in agreement with those of BHW12 at energies
$\lesssim 3\times 10^4$~GeV and $\lesssim 3\times 10^5$~GeV, respectively.
Above these energies, where the neutrino spectrum is mainly determined by the kaon decays,
we find deviations from the BHW12 results, which become
prominent mainly in the electron neutrino spectra.
The main reason for these
differences is that the DMPR12 code takes into account the
decay of all types of kaons, such as the short-lived ($K^0_{\rm S}$) and long-lived ($K^0_{\rm L}$) neutral kaons, whereas
BHW12 considered only the leading mode of $K^{+}$ production.

Inclusion of the $K^0$ decays leads not only to an absolute inrease of the electron and muon neutrino fluxes but
also to a relative increase of the electron to muon neutrino fluxes.
On the one hand, the most probable decay channel for $K^0_{\rm L}$
is $K^0_{\rm L} \rightarrow \pi^{\pm} + e^{\pm} + \nu_{\rm e}$ with a branching ratio (b.r.) of $\sim 0.45$, followed then
by the $K^0_{\rm L} \rightarrow \pi^{\pm} + \mu^{\pm} + \nu_{\mu}$ channel with a b.r. of $0.27$ \citep{beringer12}.
On the other hand, $K^0_{\rm S}$ decays mainly through the hadronic modes, i.e. 
$K^0_{\rm S} \rightarrow \pi^{+} + \pi^{-}$ (b.r.=0.69) and $K^0_{\rm S} \rightarrow \pi^{0} + \pi^0$ (b.r.=0.30) \citep{beringer12},
and thus, is not responsible for the relative increase of electron over muon neutrino flux.
Given that the DMPR12 code is written in such
a way that does not allow us to isolate the production of neutral kaons, we
cannot neglect the neutrinos produced by their decay.
However, we can calculate the neutrino spectra by taking into account
the production of only $K^{+}$ or $K^{-}$. Our results for the former case are shown with dotted lines
in Fig.~\ref{fig1-app}. 
We find that the muon neutrino flux at high energies decreases and
approaches the results of BHW12. 
It does not become identical though, because BHW12 have also 
taken into account the muon polarization in the chain of 
decaying kaons and pions which, in the case of $K^+$ with a power spectrum 
$\varpropto E^{-2}$, leads to a suppression of the muon neutrino flux by a factor of 
about $25\%$ \citep{lipari07}. 
The high-energy bump of the
electron neutrino spectrum  remains, however, practically unaltered. This demonstrates that 
it is the $K^0$ decay channel that  mainly contributes to these energies.
As a second step, we include processes (i)-(v) in the DMPR12 code in order to
investigate their effect on the neutrino spectrum.

\begin{figure}
 \centering
\includegraphics[width=0.5\textwidth, height=7cm]{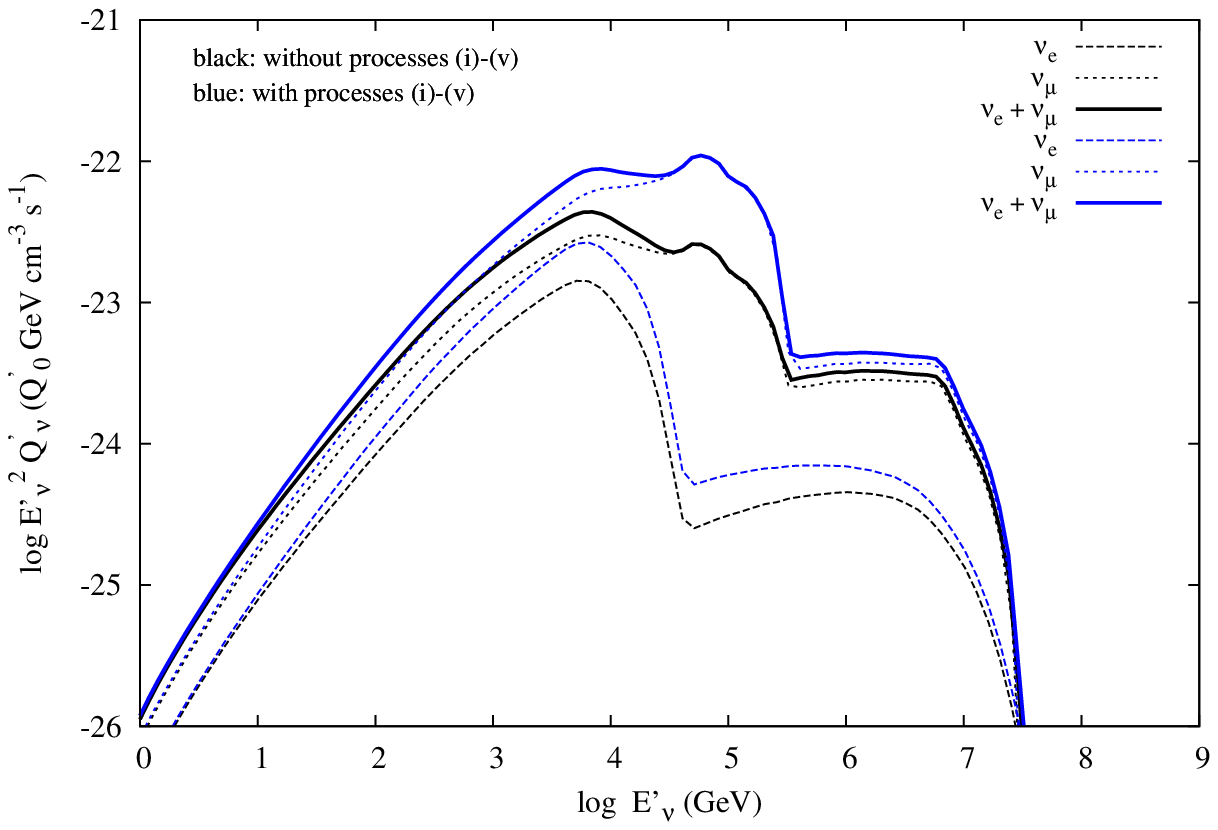}
\caption[]{Comparison of neutrino spectra calculated by including processes (i)-(v)
(blue lines) and by neglecting them (black lines). All parameters are the same as in Fig.~\ref{fig1-app}.
 \label{fig2-app}}
\end{figure}

The neutrino spectra obtained when all processes are taken into account
are shown in Fig.~\ref{fig2-app} (blue lines).
For comparison reasons, we overplotted the neutrino flux shown in Fig.~\ref{fig1-app} (black lines).
The peak flux of the total ($\nu_{\rm e}+\nu_{\mu}$) spectrum
increases at most by a factor of 4 when all processes are included. 
Although the shape of the electron neutrino spectrum is unaffected, Fig.~\ref{fig2-app}
demonstrates that inclusion of all processes enhances the peak flux of the muon neutrino spectrum 
originating from direct pion decay (at $\sim 6\times 10^4$~GeV). This is further reflected to the 
total neutrino spectrum, which becomes harder, i.e. peaking at higher energies.
Plugging into eq.~(\ref{tpg}) the parameter values used here, 
we find  $f_\pi \ll 1$, i.e. our study case
is optically thin to photopion interactions. This 
suggests that $n\gamma$ interactions do not significantly affect the neutrino spectra.
We verified that among all processes examined here, it is
the injection of secondaries, which
produce more target-photons through synchrotron radiation and/or inverse Compton scattering,
that modifies at most the neutrino spectra.
The effects of such additional processes on the neutrino spectra can be treated only by PDE solver codes,
and requires a wider search of the parameter space.

\end{document}